\documentclass{article}

\usepackage{amsmath}
\usepackage{amssymb}
\usepackage{color}
\usepackage{fullpage}
\usepackage{graphicx}
\usepackage[pdftex,colorlinks]{hyperref}
\usepackage{fourier}
\usepackage{comment}
\usepackage[T1]{fontenc}
\usepackage{frcursive}
\usepackage{amsthm}

\begin{document}
	\title{\sc Some non-relativistic implications of hydrogenic solutions to angular momenta}
	\author{\huge{ARJAN SINGH PUNIANI}}
	\date{April 2010}

\maketitle
The intention of this paper is to provide solutions to commutative relations relevant to calculations regarding the hydrogen atom (or similar monoelectronic systems). Though exact solutions exist to these systmes, the value to approximation methods stems from the ability to conveniently parse physical non-sense by comparison with these identities. A derivation accompanies each identity. 
\section{Identities} 
Since
$$
\vec{L} = \vec{r} \times \vec{p}
$$
Then
$$
\left| \begin{array}{ccc} L_x & L_y &L_z \\ p_x & p_y &p_z \\  x & y &z \\ \end{array} \right| 
$$
Thus:
$$
L_x =  yp_z - zp_y 
$$
$$
L_y = -xp_z + zp_x
$$
$$
L_z = xp_y - yp_x
$$

\subsection{ $[ L_x, x ]$}

$$
[ L_x, x ] = L_xx - xL_x = (yp_z - zp_y)x - x(yp_z - zp_y) 
$$
Apply a test function:
$$
[ L_x, x ]\Psi = (yp_z - zp_y)x\Psi - x(yp_z - zp_y)\Psi
\Rightarrow
(-i\hbar)\left[ y\frac{\partial}{\partial z}(x\Psi) - z\frac{\partial}{\partial y}(x\Psi) \right] -
(-i\hbar)\left[ xy\frac{\partial}{\partial z}(\Psi) - xz\frac{\partial}{\partial y}(\Psi) \right] 
$$
No partial derivatives are affecting $x$, thus:
$$
(-i\hbar)\left[ xy\frac{\partial}{\partial z}(\Psi) - zx\frac{\partial}{\partial y}(\Psi) \right] +
(i\hbar)\left[ xy\frac{\partial}{\partial z}(\Psi) - xz\frac{\partial}{\partial y}(\Psi) \right] 
$$
$x,y,z$ all commute, and, in general, $[p_j, i], [i, p_j] = 0$ for $i\neq j$.
Thus:
$$
(-i\hbar)\left[ xy\frac{\partial}{\partial z}(\Psi) - zx\frac{\partial}{\partial y}(\Psi) \right] +
(i\hbar)\left[ xy\frac{\partial}{\partial z}(\Psi) - xz\frac{\partial}{\partial y}(\Psi) \right] = 
(yxp_z - zxp_y) - (xyp_z - zxp_y) = 0
$$
Thus,
$$
[ L_x, x ] = 0
$$

\subsection{ $[ L_x, y ]$}
$$
[ L_x, y ] = L_xy - yL_x = (yp_z - zp_y)y - y(yp_z - zp_y) 
$$
Apply a test function:
$$
[ L_x, y ]\Psi = (yp_z - zp_y)y\Psi - y(yp_z - zp_y)\Psi
\Rightarrow
(-i\hbar)\left[ y\frac{\partial}{\partial z}(y\Psi) - z\frac{\partial}{\partial y}(y\Psi) \right] -
(-i\hbar)\left[ y^2\frac{\partial}{\partial z}(\Psi) - yz\frac{\partial}{\partial y}(\Psi) \right] 
$$
$$
(-i\hbar)\left[y^2\frac{\partial}{\partial z}(\Psi) - z\left( y\frac{\partial \Psi}{\partial y} + \Psi \right) \right] -
(-i\hbar)\left[ y^2\frac{\partial}{\partial z}(\Psi) - yz\frac{\partial}{\partial y}(\Psi) \right] =
(-i\hbar) \left[y^2\frac{\partial \Psi}{\partial z}  -zy\frac{\partial \Psi}{\partial y} -z\Psi - y^2 \frac{\partial \Psi}{\partial z} + yz\frac{\partial \Psi}{\partial y}
\right] 
$$
$$
(-i \hbar) \left( -z\Psi  \right) =  [ L_x, y ]\Psi 
$$
Thus,
$$
[ L_x, y ] = z i \hbar
$$

\subsection{ $[ L_x, z ]$}
$$
[ L_x, z ] = L_xz - zL_x = (yp_z - zp_y)z - z(yp_z - zp_y) 
$$
Apply a test function:
$$
[ L_x, z ]\Psi = (yp_z - zp_y)z\Psi - z(yp_z - zp_y)\Psi
\Rightarrow
(-i\hbar)\left[ y\frac{\partial}{\partial z}(z\Psi) - z\frac{\partial}{\partial y}(z\Psi) \right] -
(-i\hbar)\left[ zy\frac{\partial}{\partial z}(\Psi) - z^2\frac{\partial}{\partial y}(\Psi) \right] 
$$
$$
(-i\hbar)\left[y \left( z\frac{\partial \Psi}{\partial z} + \Psi \right)  - z^2\left( \frac{\partial \Psi}{\partial y}  \right) \right] -
(-i\hbar)\left[ zy \frac{\partial}{\partial z}(\Psi) - z^2\frac{\partial}{\partial y}(\Psi) \right] =
(-i\hbar) \left[zy\frac{\partial \Psi}{\partial z}  -z^2\frac{\partial \Psi}{\partial y} + y\Psi - +z^2 \frac{\partial \Psi}{\partial y} - yz\frac{\partial \Psi}{\partial z}
\right] 
$$
$$
(-i \hbar) \left( y\Psi  \right) =  [ L_x, y ]\Psi 
$$
Thus,
$$
[ L_x, z ] = -y i \hbar
$$

\subsection{$[L_x, p_x]$}
$$
L_xp_x\Psi - p_xL_x\Psi
$$
$$
(yp_zp_x - zp_yp_x)\Psi - 
(p_xyp_z - p_xzp_y)\Psi 
$$
Term-by-term:
$$
yp_zp_x\Psi = (-i\hbar)^2 y \frac{\partial^2 \Psi}{\partial z \partial x}
$$
$$
-zp_yp_x\Psi =  - (-i\hbar)^2 z \frac{\partial^2 \Psi}{\partial y \partial x}
$$
$$
p_xyp_z\Psi = (-i\hbar)^2 y \frac{\partial^2 \Psi}{\partial y \partial x}
$$
$$
- p_xzp_y\Psi = - (-i\hbar)^2 z \frac{\partial^2 \Psi}{\partial x \partial y}
$$
By Clairaut's Theorem, the order of differentiation is immaterial; thus:
$$
(-i\hbar)^2 y \frac{\partial^2 \Psi}{\partial z \partial x}  - (-i\hbar)^2 z \frac{\partial^2 \Psi}{\partial y \partial x}
- (-i\hbar)^2 y \frac{\partial^2 \Psi}{\partial y \partial x} + (-i\hbar)^2 z \frac{\partial^2 \Psi}{\partial x \partial y} = 0
$$

\subsection{ $  [L_x, p_y]  $}
$$
L_xp_y\Psi - p_yL_x\Psi \Rightarrow 
(yp_zp_y - zp_yp_y)\Psi - 
(p_yyp_z - p_yzp_y)\Psi 
$$
$$
yp_zp_y \Psi  = (-i\hbar)^2 y \frac{\partial^2 \Psi}{\partial z \partial y}
$$
$$
- zp_yp_y\Psi = - (-i\hbar)^2 z \frac{\partial^2 \Psi}{\partial y^2}
$$
$$
p_yyp_z\Psi = (-i\hbar) \left[ p_z + y \frac{\partial p_z}{\partial y} \right]\Psi
$$
$$
- p_yzp_y\Psi = - (-i\hbar)^2 z \frac{\partial^2 \Psi}{\partial y^2}
$$
Or:
$$
(-i\hbar)^2 y \frac{\partial^2 \Psi}{\partial z \partial y} - (-i\hbar)^2 z \frac{\partial^2 \Psi}{\partial y^2} - (-i\hbar) \left[ p_z + y \frac{\partial p_z}{\partial y} \right]\Psi  + (-i\hbar)^2 z \frac{\partial^2 \Psi}{\partial y^2}
$$
$$
- (-i\hbar)p_z\Psi = - (-i\hbar)^2 \frac{\partial \Psi}{\partial z} \Rightarrow [L_x, p_y ] = (i\hbar)p_z
$$

\subsection{ $  [L_x, p_z]  $}
$$
L_xp_z\Psi - p_zL_x\Psi \Rightarrow 
(yp_zp_z- zp_yp_z)\Psi - 
(p_zyp_z - p_zzp_y)\Psi 
$$
$$
yp_zp_z \Psi = (-i\hbar)^2 y \frac{\partial^2 \Psi}{\partial z^2}
$$
$$
- zp_yp_z \Psi = - (-i\hbar)^2 z \frac{\partial^2 \Psi}{\partial y \partial z}
$$
$$
p_zyp_z  \Psi = (-i\hbar)^2 y \frac{\partial^2 \Psi}{\partial z^2}
$$
$$
- p_zzp_y \Psi = (-i\hbar)  \left[ p_y + z \frac{\partial p_y}{\partial z} \right]\Psi
$$
The first and third terms cancel (after distributing the negation, and we are left with:
$$
- (-i\hbar)^2 z \frac{\partial^2 \Psi}{\partial y \partial z} - (-i\hbar) p_y\Psi  +  (-i\hbar)^2 z \frac{\partial^2 \Psi}{\partial y \partial z}
$$
Thus:
$$
[L_x, p_z] \Psi = - (-i\hbar) p_y\Psi \Rightarrow [L_x, p_z] = (i\hbar) p_y
$$

\subsection{ $  [L_x, L_y]  $}
$$
(yp_z - zp_y)(-xp_z + zp_x) - (-xp_z + zp_x)(yp_z - zp_y) 
$$

$$
yp_zzp_x - yp_zxp_z - zp_yzp_x + zp_yxp_z -
(zp_xyp_z - zp_xzp_y - xp_zyp_z + xp_zzp_y) = 
$$
$$
yp_zzp_x - yp_zxp_z - zp_yzp_x + zp_yxp_z - zp_xyp_z + zp_xzp_y + xp_zyp_z - xp_zzp_y
$$
We now know which terms commute, so collect the ones that do not:
$$
[yp_zzp_x - xp_zzp_y]\Psi
$$
$$
y(-i\hbar) \left[  z \frac{\partial p_x}{\partial z} + p_x      \right]\Psi - 
x(-i\hbar)   \left[  z \frac{\partial p_y}{\partial z} + p_y     \right]\Psi
$$
$$
yz(-i\hbar)^2 \frac{\partial^2 \Psi}{\partial x \partial z} + y(-i\hbar)p_x\Psi - 
xz(-i\hbar)^2 \frac{\partial^2 \Psi}{\partial z \partial y} -  y(-i\hbar)p_y\Psi 
$$
Note that
$$
[A-B,C-D] = [A,C] - [D,A] - [B,C] + [B,D] 
$$
So 
$$
[yp_z - zp_y,zp_x -xp_z ]
$$
Is 
$$
 [yp_z, zp_x ] - [yp_z, xp_z] - [zp_y, zp_x] + [zp_y, xp_z]
$$
And by similar argument, the middle terms will eventually vanish towards the end of the calculation, so we neglect them now:
$$
 [yp_z, zp_x ]  +  [zp_y, xp_z] = yp_x[z,p_z] + xp_y[ z,p_z] = i\hbar(xp_y-yp_x) = i\hbar L_z
$$
Since
$$
yp_xzp_z - yp_xp_zz+ xp_yzp_z - xp_yp_zz
$$

\subsection{ $  [L_x, \vec{L}^2]  $}
$$
 [L_x, L_x^2] +  [L_x, L_y^2] +  [L_x, L_z^2]
$$
$$
 [L_x, L_x^2] = 0 
$$
So
$$
[L_x, L_y]L_y + L_y[L_x, L_y]L_y + [L_x, L_z]L_z + L_z[L_x, L_z]
$$
$$
(-i\hbar L_z)L_y + L_y(-i\hbar L_z) + (-i\hbar L_y)L_z + L_z(-i\hbar L_y)
$$

Since
$$
[AB,C] = A[B,C] + [A,C]B = ABC-ACB + ACB - CAB = ABC - CAB = [AB,C] 
$$
Then:
$$
(-i\hbar L_z)L_y + L_y(-i\hbar L_z) + (-i\hbar L_y)L_z + L_z(-i\hbar L_y) = 0
$$

\subsection{ $  [L_x, \vec{r}^2]  $}
$$
 [L_x, x^2] +  [L_x, y^2] +  [L_x, z^2]
$$
$$
 [L_x, x^2] = L_xx^2 - x^2L_x = 0
$$
So
$$
L_xy^2 - y^2L_x + L_xz^2 - z^2L_x
$$

\subsection{ $ [L_x, \vec{p}^2]  $}
$$
 [L_x, p_x^2] +  [L_x, p_y^2] +  [L_x, p_z^2]
$$
$$
 [L_x, L_x^2] = 0 
$$
So

\section{Theorem: Skew Hermitian operators have pure imaginary eigenvalues}
\begin{proof}
If 
$$
\langle \hat{\Omega} \Psi | \Psi \rangle  =  - \langle \Psi |  \hat{\Omega}  \Psi \rangle
$$
And
$$
\langle \hat{\Omega} \Psi | \Psi \rangle  = \Omega^* \langle \Psi | \Psi \rangle
$$
$$
\langle \Psi |  \hat{\Omega}  \Psi \rangle = \langle \Psi |   \Psi \rangle \Omega
$$
Then, by definition, the skew-hermitian operator $\hat{\Omega}$ satisfies:
$$
\Omega^* \langle \Psi | \Psi \rangle = - \left( \langle \Psi |   \Psi \rangle \Omega \right) \Rightarrow 
\Omega^* = -\Omega
$$
If $\Omega \in \mathbb{C}$, then it assumes the form $A + iB$ where $A,B \in \mathbb{R}$ and
$$
\Omega^* = -\Omega \Rightarrow (A + iB)^* = -(A + iB) 
$$
Or
$$
A - iB =  -A - iB
$$
Which can only be true if $A = 0$. Thus, the observables (eigenvalues) from skew Hermitian operators are always purely imaginary. 
\end{proof}
\section{References}
\begin{enumerate}
\item Boas, Mary \emph{Mathematical Methods for the Physical Sciences} Wiley, Sons, Co. 2002
\item Griffiths, D.J. \emph{Introduction to Quantum Mechanics} Harper-Collins 2003
\end{enumerate}
\end{document}